\documentclass[prl,showpacs,twocolumn,aps]{revtex4}
\usepackage{graphicx}
\newcommand{\be}{\begin{equation}}
\newcommand{\ee}{\end{equation}}
\newcommand{\bea}{\begin{eqnarray}}
\newcommand{\eea}{\end{eqnarray}}
\newcommand{\up}{\uparrow}
\newcommand{\dn}{\downarrow}
\newcommand{\bwt}{\begin{widetext}}
\newcommand{\ewt}{\end{widetext}}
\begin{document}
\title{
Spin Torque and its Relation to Spin Filtering
}
\author{Wonkee Kim and F. Marsiglio}
\affiliation{
Department of Physics, University of Alberta, Edmonton, Alberta,
Canada, T6G~2J1}
\begin{abstract}
The spin torque exerted on a magnetic moment is a reaction to spin filtering when 
spin-polarized electrons interact with a thin ferromagnetic film. We show that, for certain
conditions, a spin transmission resonance (STR) 
gives rise to a failure of spin filtering. As a consequence,
no spin is transfered to the ferromagnet. 
The condition for STR depends
on the incoming energy of electrons and the thickness of the film.
For a simple model we find that when the STR condition is satisfied, the ferromagnetic
film is transparent to the incoming electrons.
\end{abstract}

\pacs{75.70.Ak,72.25.-b,85.75.-d}
\date{\today}
\maketitle

Since the spin torque problem was conceptualized\cite{slonczewski,berger} 
and observed experimentally\cite{tsoi1,myers,katine,wegrowe}, 
enormous attention has been paid
to the spin torque of a magnetic moment, driven by a spin-polarized current,
both theoretically\cite{bazaliy,sun,waintal,stile,zhang,kim} 
and experimentally\cite{tsoi2,albert,ji,grollier,kiselev}.
These discoveries also
open a door to new magnetic devices and applications to, for example,
magnetic random access memory\cite{myers}, spin-wave amplification
by stimulated emission of radiation\cite{tsoi2}, and
nanoscale microwave sources\cite{kiselev}.
However, it is still necessary to understand and explore the physics underlying
this phenomenon. In particular, the significance of the thickness of the ferromagnetic
film has not been fully studied at present.

The key interaction associated with the spin torque is
an interaction between the incoming spins $({\bf s})$
and the magnetic moment $({\bf M})$:
$-2J_{H}{\bf s}\cdot{\bf M}$, where $J_{H}$ is the coupling strength.
Thus, the Hamiltonian for this problem has two ingredients;
one is the kinetic energy of the incoming electrons and the other
the interaction energy with the moment in the film. The magnetic moment is assumed to
originate from the local spins in the ferromagnet and its magnitude
$M_{0}$ is a constant.

As we illustrate in what follows, the spin torque arises as a reaction to spin filtering.
Suppose a spin polarized electron beam enters normally to
the ferromagnetic film. After spin filtering, incoming electrons
lose their spin components perpendicular to the magnetic moment.
Because of conservation laws, this spin is transferred as a 
torque exerting on the moment. Interestingly, such a mechanism 
for `spin transfer' implies that no spin torque will take place if the
incoming spin is not filtered. In this paper we scrutinize a simple model
to understand what conditions are necessary for zero spin transfer and
concomitant absence of spin filtering.

No spin will be transfered from the incoming spins 
to the magnetic moment when the condition for a spin transmission resonance (STR)
is satisfied.
STR is a purely quantum mechanical phenomenon and is similar to
the well known particle transmission resonance (PTR) \cite{schiff} 
for a potential barrier. In our case the spin state of incoming electrons remains
unaltered even after interacting with the magnetic moment in the ferromagnet
and, thus, no spin filtering occurs.
Since the spin torque is a reaction to spin filtering,
the magnetic moment does not experience a spin torque.
For a given energy of incoming electrons, 
this phenomenon is periodic with a thickness depending 
only on the interaction energy and fundamental constants like the electron mass.
STR is one of the unique characteristics of spin transfer from a current to a
moment, and can distinguish this mechanism unambiguously from that obtained through an 
applied magnetic field (in this case the field would be induced by the applied current).

When a spin torque is exerted on the magnetic moment, it will align 
parallel to the direction of incoming spins after a relaxation time
$\tau_{0}$.
Thus, a signature of zero spin transfer is that $\tau_{0}\rightarrow\infty$.
Using the adiabatic 
approximation, we derive the equations of motion
for the magnetic moment and obtain an analytic form of $\tau_{0}$.
The adiabatic approximation is valid if $\tau_{0}$ is much longer than
a time scale incoming electrons spend on interacting with
the magnetic moment. Obviously, this approximation is always
applicable near STR.
Plots of $\tau_{0}$ as a function of the thickness of the ferromagnetic film
and the energy of incoming electrons illustrate the physics associated with STR.
In the derivation of the equations of motion for the magnetic moment, it is necessary
to obtain the electron wave function inside the film to evaluate the expectation
value of the spin operator. For simplicity, we consider a single-domain
ferromagnet as in Refs.\cite{slonczewski,berger,sun,waintal,stile,zhang}.

\begin{figure}[tp]
\begin{center}
\includegraphics[height=2.6in,width=3.3in]{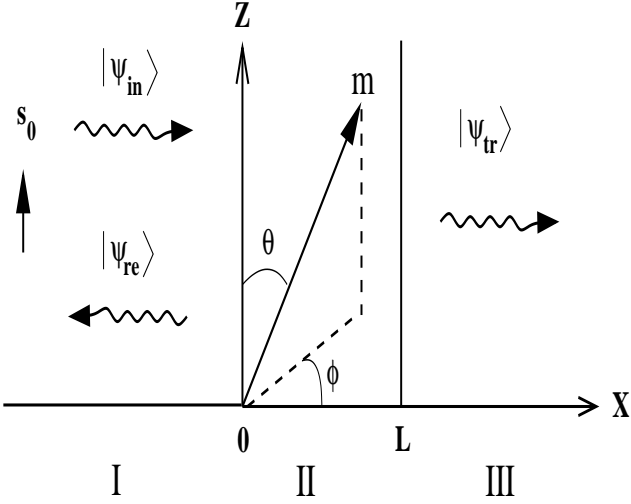}
\caption{Quantum mechanical problem associated with
the spin transfer.  Region II represents the ferromagnetic film
with the thickness $L$. Regions I and III are non-ferromagnetic.
}
\end{center}
\end{figure}

Depicted in Fig.~1 is the quantum mechanical problem we consider to obtain
the electron wave function. Region II represents the ferromagnetic film
with the thickness $L$
while regions I and III are non-ferromagnetic. Since the film is parallel to the
YZ plane, the Schr{\" o}dinger equation relevant to our problem is one-dimensional.
We suppose incoming electrons with momentum ${\bf k}$ along the X axis.
The direction of the polarized spins is chosen to be Z axis.
Then the incoming wave function $|\psi_{in}\rangle$ is $|+\rangle e^{ikx}$, 
where $|+\rangle$ is the spin-up state in the lab frame. The incoming
energy $\epsilon_{in}=(\hbar k)^{2}/2m$, where $m$ is the electron mass.
We will utilize a normalization constant ${\cal C}$ 
to obtain dimensionless equations of motion for the magnetic moment\cite{kim}. 
The direction of the magnetic moment is defined by $\theta$ and $\phi$
inside the ferromagnetic film. For a given direction of the magnetic
moment, we solve the Schr{\" o}dinger equation to obtain the electron wave 
function. 

In region I,
the wave function consists of the incoming $|\psi_{in}\rangle$
and reflected $|\psi_{re}\rangle$
wave function while in region III, there is only the transmitted 
$|\psi_{tr}\rangle$ wave function.
The reflected 
and transmitted wave functions are expressed 
in terms of eigenstates
$|\chi_{\sigma}\rangle$ of the
interaction $2J_{H}{\bf M}\cdot{\bf s}$ such that
$2J_{H}{\bf M}\cdot{\bf s}|\chi_{\sigma}\rangle=
\pm J_{H}M_{0}|\chi_{\sigma}\rangle$, where $\sigma=\up$ or $\dn$,
and $+(-)$ is for $\sigma=\up(\dn)$.
We assume that the ferromagnetic film is sufficiently clean
that the mean free path of electrons within the film is much longer
than the film thickness. Therefore the convergence factor introduced in
Refs.\cite{berger,kim} is not necessary, and we can investigate 
the significance of the thickness explicitly. 
Because of the interaction, the momentum is spin split;
$k_{\sigma}=\sqrt{k\pm 2mJ_{H}M_{0}}$, depending on the eigenstates 
$|\chi_{\sigma}\rangle$.
If $\epsilon_{in}$ is less than
$J_{H}M_{0}$, $k_{\dn}=i\kappa_{\dn}$ becomes pure imaginary where
$\kappa_{\dn}=\sqrt{2mJ_{H}M_{0}-k}$ and the corresponding wave function
decays exponentially. In this case, however, STR will not occur, as one might
expect based on the usual PTR conditions.
Now the wave functions in regions I, II, and III can be written as follows:
\bwt
\bea
|\psi_{I}\rangle&=&|+\rangle e^{ikx}
+\Bigl[
R_{\up}|\chi_{\up}\rangle\langle\chi_{\up}|+\rangle+
R_{\dn}|\chi_{\dn}\rangle\langle\chi_{\dn}|+\rangle\Bigr] e^{-ikx}
\nonumber\\
|\psi_{II}\rangle&=&\left(A_{\up}e^{ik_{\up}x}+B_{\up}e^{-ik_{\up}x}\right)
|\chi_{\up}\rangle\langle\chi_{\up}|+\rangle
+\left(A_{\dn}e^{ik_{\dn}x}+B_{\dn}e^{-ik_{\dn}x}\right)
|\chi_{\dn}\rangle\langle\chi_{\dn}|+\rangle
\nonumber\\
|\psi_{III}\rangle&=&
\Bigl[
T_{\up}|\chi_{\up}\rangle\langle\chi_{\up}|+\rangle+
T_{\dn}|\chi_{\dn}\rangle\langle\chi_{\dn}|+\rangle\Bigr] e^{ikx}.
\eea
\ewt

The coefficients $R_{\sigma}$, $A_{\sigma}$, $B_{\sigma}$, and $T_{\sigma}$
are determined by the boundary conditions of wave functions at $x=0$ 
and $x=L$; namely,
$\langle\pm|\psi_{I}(0)\rangle=\langle\pm|\psi_{II}(0)\rangle$,
$\langle\pm|\psi_{II}(L)\rangle=\langle\pm|\psi_{III}(L)\rangle$,
and similar relations for their derivatives. Some straightforward algebra yields
\bea
R_{\sigma}&=&\frac
{(k^{2}-k^{2}_{\sigma})\left(1-e^{2ik_{\sigma}L}\right)}
{(k+k_{\sigma})^{2}-(k-k_{\sigma})^{2}e^{2ik_{\sigma}L}}
\nonumber\\
A_{\sigma}&=&\frac
{2k(k_{\sigma}+k)}
{(k+k_{\sigma})^{2}-(k-k_{\sigma})^{2}e^{2ik_{\sigma}L}}
\nonumber\\
B_{\sigma}&=&\frac
{2k(k_{\sigma}-k)e^{2ik_{\sigma}L}}
{(k+k_{\sigma})^{2}-(k-k_{\sigma})^{2}e^{2ik_{\sigma}L}}
\nonumber\\
T_{\sigma}&=&\frac
{4kk_{\sigma}e^{i(k_{\sigma}-k)L}}
{(k+k_{\sigma})^{2}-(k-k_{\sigma})^{2}e^{2ik_{\sigma}L}}.
\eea
It is worthwhile noting the similarity between our case and the
one-dimensional potential barrier problem. At a glance, one can
see the above coefficients are exactly the same as those for the
potential barrier except for their two-fold nature due to the
spin index. In fact, the two-fold nature can be mapped onto
a potential {\it well} for $\sigma=\up$ and a potential {\it barrier}
for $\sigma=\dn$. Note that PTR takes place in a potential
barrier as well as in a potential well. Therefore, conditions for
STR are effectively those for PTR for the corresponding barrier 
and well at the same time.

For PTR, zero reflectance guarantees a transmission resonance;
this is not the case for STR. If $k_{\sigma}L=n_{\sigma}\pi$
$(n_{\sigma}=1,\;2,\;3,\;\cdots)$, then $R_{\sigma}=0$. However,
STR does not occur yet because $T_{\sigma}=e^{i(k_{\sigma}-k)L}$;
in other words, it is not guaranteed that 
$\langle-|\psi_{III}\rangle=0$, which means the spin state in region III
can not be represented only by $|+\rangle$. Since the incoming wave
function has only $|+\rangle$, non-zero $\langle-|\psi_{III}\rangle$ indicates
that spin is transfered to the magnetic moment. The condition for STR
is $k_{\sigma}L=(2n_{\sigma}-1)\pi$ or $2n_{\sigma}\pi$ with $n_{\dn}<n_{\up}$
due to $k_{\dn} < k_{\up}$.
If the above condition is satisfied, we obtain
$\langle+|\psi_{III}\rangle=e^{ik(x-L)}$ and $\langle-|\psi_{III}\rangle=0$.
This means that the transmitted wave function remains
unaltered even after interacting with the magnetic moment in the 
ferromagnetic thin film except for an additional phase $e^{-ikL}$
depending only on the thickness. Consequently, spin filtering fails
completely under this condition.

Let us examine the condition for STR in more detail.
Suppose $2n_{\sigma}\pi$ and the incoming energy is larger than
the interaction energy $J_{H}M_{0}$ by a factor of $\eta$; 
$\epsilon_{in}=\eta J_{H}M_{0}$, and $k_{\up}=\sqrt{\eta+1}k_{0}L$
and $k_{\dn}=\sqrt{\eta-1}k_{0}L$, where $k_{0}=\sqrt{2mJ_{H}M_{0}}$.
We now obtain constraints for $\eta$ and $L$ under the
STR condition:
$\eta=\left[1+(n_{\dn}/n_{\up})^{2}\right]/\left[1-(n_{\dn}/n_{\up})^{2}\right]$
and $L=n_{\up}\sqrt{2\left[1-(n_{\dn}/n_{\up})^{2}\right]}L_{0}$, where 
$L_{0}=\pi/k_{0}$. For $k_{\sigma}L=(2n_{\sigma}-1)\pi$, a similar constraint
can be obtained by replacing $n_{\sigma}$ with $n_{\sigma}-1/2$.
This analysis tells us that the STR condition can be satisfied by controlling
$\eta$ and $L$; for example, if $\eta=5/4$ and $L=2L_{0}$, then STR takes place.
Another example is for $\eta=5/3$ and $L=\sqrt{6}L_{0}$. A more interesting
result of the analysis is that for a given value of $\eta$, the thickness $L$
is periodic. This arises because $\eta$ is determined by the ratio
$n_{\dn}/n_{\up}$ while $L$ is proportional to $n_{\up}$. Therefore,
for a given value of $\epsilon_{in}$, say, $\eta=5/3$, STR will take place
when $L=\sqrt{6}L_{0},\;2\sqrt{6}L_{0},\;3\sqrt{6}L_{0},\;\cdots$.
This is inherently a quantum mechanical property and 
completely different from the effects of a current-induced magnetic 
field on the magnetic moment.

Using the wave function in region II, $|\psi_{II}\rangle$,
we evaluate the expectation value of the spin operators to
derive equations of motion for the magnetic moment for an arbitrary
case. If the STR condition is met, we will be able to see its signature in the
equations. Since the spin expectation is evaluated by
$\langle s_{i}\rangle=(1/2)\langle\psi_{II}|\sigma_{i}|\psi_{II}\rangle$,
where $\sigma_{i}$ $(i=x,\;y,\;z)$ are Pauli matrices, we obtain
\bea
\langle s_{x}\rangle&=&\alpha m_{x}-\beta m_{y}-\gamma m_{z}m_{x}
\nonumber\\
\langle s_{y}\rangle&=&\alpha m_{y}+\beta m_{x}-\gamma m_{z}m_{y}
\nonumber\\
\langle s_{z}\rangle&=&\alpha m_{z}+\gamma\left(1-m^{2}_{z}\right)
\eea
where 
\bea
\alpha&=&\frac{1}{4}\int^{L}_{0}dx\;\Bigl[|C_{\up}|^{2}(1+m_{z})
-|C_{\dn}|^{2}(1-m_{z})\Bigr]
\nonumber\\
\beta&=&\frac{1}{2}\int^{L}_{0}dx\;\mbox{Im}\left[C^{*}_{\dn}C_{\up}\right]
\nonumber\\
\gamma&=&\frac{1}{2}\int^{L}_{0}dx\;\mbox{Re}\left[C^{*}_{\dn}C_{\up}\right]
\eea
with $C_{\sigma}=A_{\sigma}e^{ik_{\sigma}x}+B_{\sigma}e^{-ik_{\sigma}x}$.
Note that we also take an average over region II to take into account
a net effect of the incoming spins on the magnetic moment as done
in Refs.\cite{berger,zhang,kim}.

The equation of motion for the magnetic moment corresponding to
the interaction is ${d{\bf M}}/{d t}=
2\gamma_{0}J_{H}{\bf M}\times
\langle {\bf s}\rangle$, where $\gamma_{0}$ is the gyromagnetic ratio.
In order to obtain dimensionless equations we need to restore
the normalization constant ${\cal C}$, with which
the intensity of incoming beam can be controlled because 
$|{\cal C}|^{2}$ represents
the number $(N_{e})$ of incoming electrons per unit length. 
We also introduce an effective local spin $S_{local}$ to
semiclassically describe
$M_{0}=\gamma_{0}S_{local}$.
A dimensionless time
$\tau$ can be defined as $\tau=j_{0}t$, where 
$j_{0}=(\pi N_{e}/S_{local})k_{0}/m$ 
is the one-dimensional current density with 
$(\pi N_{e}/S_{local})$ electrons per unit length.
Then, we obtain
${d{\bf m}}/{d\tau}={\bf m}
\times\langle{\bf s}\rangle$:
\bea
\frac{d m_{x}}{d \tau}&=&-{\beta}m_{z}m_{x}+{\gamma}m_{y}
\nonumber\\
\frac{d m_{y}}{d \tau}&=&-{\beta}m_{z}m_{y}-{\gamma}m_{x}
\nonumber\\
\frac{d m_{z}}{d \tau}&=&{\beta}\left(1-m^{2}_{z}\right)\;.
\label{eom}
\eea
Note that the equation for $m_{z}$ does not depend on other components
of ${\bf m}$.
This equation can be solved analytically and we obtain
\be
m_{z}(\tau)=\tanh\Bigl[{\beta}\tau+\frac{1}{2}
\ln\left(\frac{1+m_{0}}{1-m_{0}}\right)\Bigr]
\ee
where $m_{0}$ is the initial value of $m_{z}$ at $\tau=0$
and $|m_{0}|<1$. 
We found 
for $\eta\le1$, ${\beta}$ is positive definite. However,
for $\eta>1$, ${\beta}\ge0$ depending on $\eta$ and the thickness $L$.
When STR occurs, it can be shown analytically that ${\beta}=0$.
Then one may assume that the magnetic moment will precess with 
a frequency ${\gamma}$ based on Eq.~(\ref{eom}); however,
one can show that ${\gamma}$ also vanishes under the condition
for STR. Consequently, the magnetic moment remains a constant in this case.

If ${\beta}>0$, we obtain an asymptotic expression for $m_{z}(\tau)$:
\be
m_{z}(\tau)\simeq1-\left(\frac{1-m_{0}}{1+m_{0}}\right)e^{-2{\beta}\tau}
\ee
It is natural to define the relaxation time $\tau_{0}$ as $1/2{\beta}$
based on the above equation. In Fig.~2 we plot $\tau_{0}$ 
as a function
of $L$ for given values of $\eta$. 
When the condition for STR is satisfied,
$\tau_{0}\rightarrow\infty$ because ${\beta}\rightarrow0$. The
singular behavior is periodic as we analyzed earlier. The period depends on
$\eta$. For $\eta=5/4$, it is $2L_{0}$ while for $\eta=5/3$ it is $\sqrt{6}L_{0}$.
For comparison, $\eta=1$ (solid) and $\eta=3/2$ (dashed curve) are also plotted.
The singularities illustrated in Fig.~2 are strong; for example, $\tau_0 \approx
(L-2L_{0})^{-2}$ for $\eta=5/4$.
Therefore, it should be measurable as long as the ferromagnetic
film is reasonably smooth. We also plot $\tau_{0}$ as 
a function $\eta$ for given values of $L$ in Fig.~3. 
For $L=2L_{0}$ and $\sqrt{6}L_{0}$, $\tau_{0}$ increases without bound
when $\eta=5/4$ and $5/3$, respectively. For any other value of $L$,
$\tau_{0}$ is finite as shown by a solid curve $(L=L_{0})$ and
by a dashed curve $(L=2.2L_{0})$. 
This plot can be used
for experiments to test our predictions
because it does not require a set of ferromagnetic films with varying
width, as required for Fig.~2.
In our estimation, $L_{0}\sim5$nm assuming $J_{H}M_{0}$ is of order
$10^{-2}$eV ($\sim$ the s-d exchange energy\cite{mitchell}). 
A typical relaxation time can be also estimated to be
of order $10^{-9}$ second if $(\pi N_{e}/S_{local})\sim 10^{4}$/cm,
which corresponds to an electron beam with intensity of order
$10^{19}$/sec/cm$^{2}$. Incoming
electrons with $\epsilon_{in}\sim10^{-2}$eV spend only
$10^{-14}$ second per nanometer within the ferromagnetic film.
Under the adiabatic approximation, the beam intensity may not
be too large because if so, $\tau_{0}$ could be comparable
with a typical time incoming electrons spend on interacting
with the magnetic moment.
However, the adiabatic 
approximation is always applicable near STR 
regardless of the beam intensity since $\tau_{0}\rightarrow\infty$
under the STR conditions.  
For an experimental test one could use 
the experimental setup
in Refs.\cite{lassailly,oberli} or the ferromagnet-normal-metal-ferromagnet
(F$_{1}$NF$_{2}$) junction. If the STR condition is satisfied in F$_{1}$,
then F$_{1}$ is transparent to the incoming spins. Therefore,
the spin torque will appear only in F$_{2}$. 

\begin{figure}[tp]
\begin{center}
\includegraphics[height=2.6in,width=3.3in]{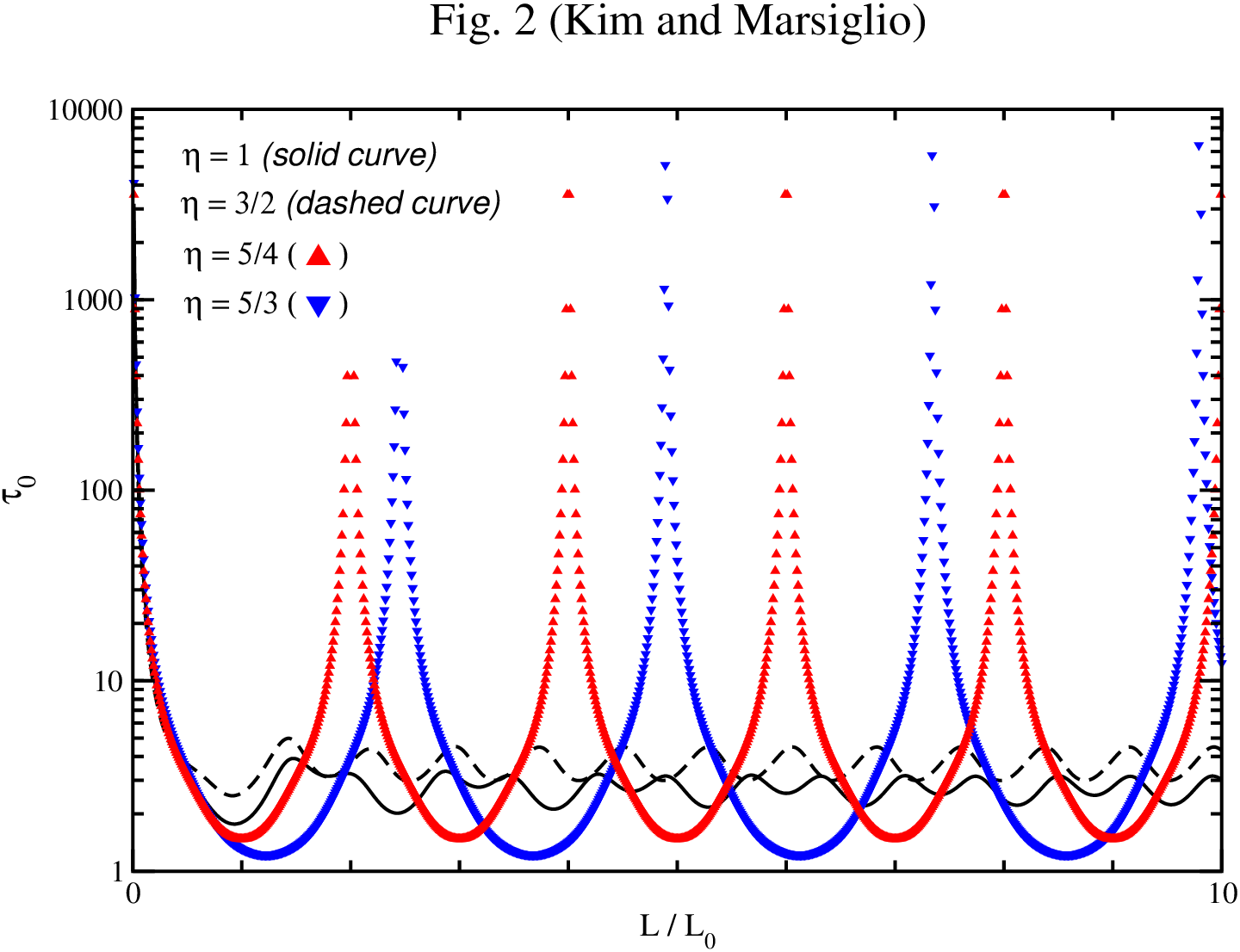}
\caption{
Relaxation time $\tau_{0}$
as a function
of $L$ for given values of $\eta$. When the condition for STR is satisfied,
$\tau_{0}\rightarrow\infty$ because ${\beta}\rightarrow0$. The
singular behavior is periodic. The period depends on
$\eta$. For $\eta=5/4$, it is $2L_{0}$ while for $\eta=5/3$ it is $\sqrt{6}L_{0}$.
For comparison, we also plot $\tau_{0}$ for $\eta=1$ (solid) and $3/2$ (dashed curve).
Note that for the estimates of parameters given in the text, a value
of $\tau_{0}\sim10$ on this plot corresponds to 
a real relaxation time $\sim10^{-9}$ second.
}
\end{center}

\begin{center}
\includegraphics[height=2.6in,width=3.3in]{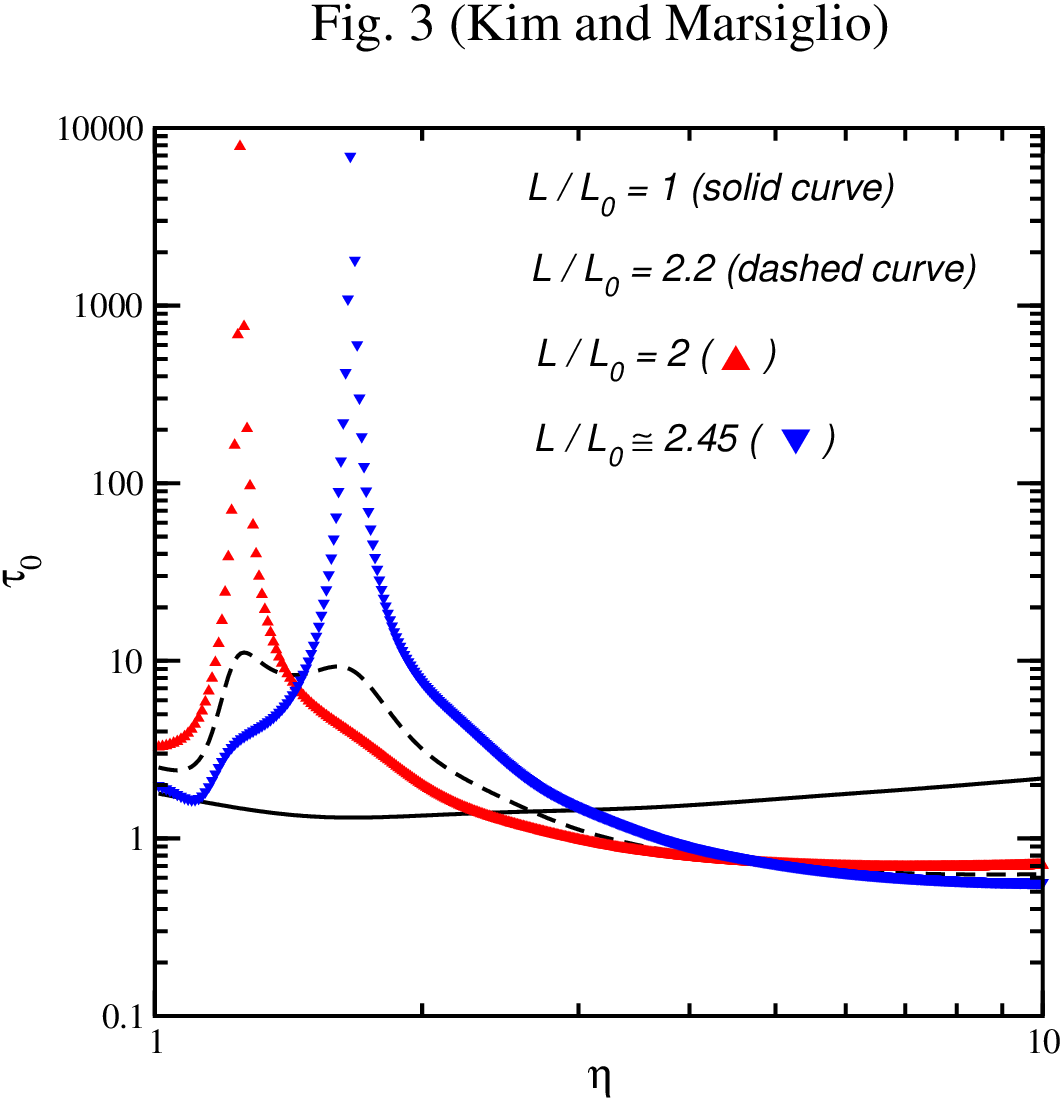}
\caption{Relaxation time $\tau_{0}$
as a function $\eta$ for given $L$.
For $L=2L_{0}$ and $\sqrt{6}L_{0}$, $\tau_{0} \rightarrow \infty$
when $\eta=5/4$ and $5/3$, respectively. For any other value of $L$,
$\tau_{0}$ is finite as shown by the solid curve $(L=L_{0})$ and
the dashed curve $(L=2.2L_{0})$.
}
\end{center}
\end{figure}

In summary, we investigated the conditions for
spin transmission resonance (STR), and their consequences. We found that
STR will occur for a variety of widths which are multiples of a
fundamental thickness (for given electron energy).
This is a quantum mechanical property and indicates
the significance of the thickness of the ferromagnetic film.
When the STR condition is satisfied,
the magnetic moment in the ferromagnetic film remains unaltered since
no spin transfer occurs.

\bibliographystyle{prl}

\end{document}